\documentclass[epj]{webofc}
\usepackage[varg]{txfonts} 
\usepackage{lineno}
\usepackage{graphicx, amsmath, color, mathrsfs}
\usepackage{epsfig}
\usepackage{epstopdf}
\graphicspath{{./figures/},{./}}
\newcommand{\beq}{\begin{equation}}
\newcommand{\eeq}{\end{equation}}
\newcommand{\bea}{\begin{eqnarray}}
\newcommand{\ena}{\end{eqnarray}}

\newcommand{\apj}{Astrophys. J.}
\newcommand{\mnras}{M.N.R.A.S.}
\begin{document}
%
%
\title{The proton and helium anomalies in the light of the Myriad model}
%
%

\author{
Pierre~Salati\,\footnote{Presentation given by P.~Salati at the 6th RICAP Conference on June 23rd, 2016.}\,\inst{1}\fnsep\thanks{\email{salati@lapth.cnrs.fr}}
\and
Yoann~G\'enolini\,\inst{1}\fnsep\thanks{\email{genolini@lapth.cnrs.fr}}
\and
Pasquale Serpico\inst{1}
\and
Richard Taillet\inst{1}}

\institute{
LAPTh, Universit\'e de Savoie Mont Blanc, CNRS, 9 Chemin de Bellevue,\\
B.P.110, Annecy-le-Vieux F-74941, France
\vskip 0.25cm
LAPTH-Conf-066/16}

\abstract{
A hardening of the proton and helium fluxes is observed above a few hundreds of GeV/nuc. The distribution of local sources of primary cosmic rays has been suggested as a potential solution to this puzzling behavior. Some authors even claim that a single source is responsible for the observed anomalies. But how probable these explanations are? To answer that question, our current description of cosmic ray Galactic propagation needs to be replaced by the Myriad model. In the former approach, sources of protons and helium nuclei are treated as a jelly continuously spread over space and time. A more accurate description is provided by the Myriad model where sources are considered as point-like events. This leads to a probabilistic derivation of the fluxes of primary species, and opens the possibility that larger-than-average values may be observed at the Earth.
For a long time though, a major obstacle has been the infinite variance associated to the probability distribution function which the fluxes follow. Several suggestions have been made to cure this problem but none is entirely satisfactory. We go a step further here and solve the infinite variance problem of the Myriad model by making use of the generalized central limit theorem. We find that primary fluxes are distributed according to a stable law with heavy tail, well-known to financial analysts. The probability that the proton and helium anomalies are sourced by local SNR can then be calculated.
The p-values associated to the CREAM measurements turn out to be small, unless somewhat unrealistic propagation parameters are assumed.
}
\maketitle
%
%
\section{Gambling with the discreteness of cosmic ray sources}
\label{sec:intro}

The proton and helium spectra are well described by a power-law distribution up to an energy of $\sim$ 350 GeV/nuc, above which a hardening is observed. This putative anomaly was reported by the PAMELA collaboration~\cite{Adriani:2011cu} and has been recently confirmed by the precision AMS-02 measurements~\cite{2015PhRvL.114q1103A,PhysRevLett.115.211101}. The proton and helium fluxes measured by the CREAM balloon borne detector~\cite{2011ApJ...728..122Y} are clearly in excess of a simple power-law behavior.

Various explanations have been proposed to account for the observed hardening, such as the existence of distinct populations of primary cosmic ray (CR) accelerators with different injection spectra~\cite{Zatsepin:2006ci}, sources characterized by a double injection spectrum like the magnetized winds of exploding Wolf-Rayet and red supergiant stars~\cite{2010ApJ...725..184B}, or the possibility of a retro-action of cosmic rays themselves on the properties of the plasma, with the consequence of a softer diffusion coefficient $K$ at high energies~\cite{Blasi:2012yr}.

Another possibility lies in the presence of local accelerators which, if considered as point-like objects, may yield a contribution somewhat different from what is expected from sources continuously spread in space and time inside our neighborhood. According to~\cite{2012MNRAS.421.1209T}, a few nearby remnants can be responsible for the observed spectral changes. In a more consistent analysis~\cite{Bernard:2012pia} where all sources, either remote or local, have the same power-law injection spectrum, the proton and helium data are well fitted with local injectors borrowed from the Green catalog and the ATNF pulsar database. There is no need to modify CR transport and even the MED CR propagation model~\cite{Donato:2003xg} yields a $\chi^{2}$ of 1.3 per dof. The fit improves considerably if the somewhat unrealistic CR model A is assumed~\cite{Bernard:2012pia}.

The question in which we are interested here is to know whether such an explanation is natural or not. In the conventional approach, sources are described by a continuous jelly in space and time. Here, the local sources are treated as point-like objects and their distribution is such that they yield a larger flux. But is this probable? To address this question and the more general problem of the stochasticity of CR fluxes yielded by injectors localized in space and time, we need the so-called Myriad model.

\section{The Myriad model: a framework for statistics}
\label{sec:myriad_model}

Once injected in the interstellar medium, primary cosmic rays propagate inside the turbulent Galactic magnetic fields. CR transport is basically understood as a diffusion process taking place within a magnetic halo which is generally pictured as a circular slab, matching the shape of the Milky Way, inside which a thin Galactic disk of gas and stars is sandwiched. The CR density $\psi = {dn}/{dE}$ is related to the CR flux $\Phi \equiv (v_{CR} / 4 \pi) \, \psi$ and fullfills the diffusion equation
\beq
{\displaystyle \frac{\partial_{\,} \psi}{\partial t}} \, - \,
K \, \Delta \psi \, = \, q_{\rm acc}(\mathbf{x},t) \; ,
\label{eq:master}
\eeq
where $q_{\rm acc}(\mathbf{x},t)$ accounts for the {\it continuous} space-time distribution of accelerators. The solution to the master equation~(\ref{eq:master}) is given by the convolution, over the magnetic halo (MH) and the past, of the CR propagator $G(\mathbf{x} , t \, \leftarrow \, \mathbf{x}_{S} , t_{S})$ with the source distribution $q_{\rm acc}(\mathbf{x}_{S},t_{S})$
\beq
\psi(\mathbf{x},t) \, = \,
{\displaystyle \int_{- \infty}^{t}} dt_{S} \,
{\displaystyle \int_{\rm MH}} d^{3}\mathbf{x}_{S} \;
G(\mathbf{x} , t \, \leftarrow \, \mathbf{x}_{S} , t_{S}) \,
q_{\rm acc}(\mathbf{x}_{S},t_{S}) \; .
\eeq
The propagator describes the probability that a CR species injected at point $\mathbf{x}_{S}$ and time $t_{S}$ diffuses at location $\mathbf{x}$ at time $t$. Assuming that $q_{\rm acc}$ does not depend in time leads to the steady state canonical solution produced by most CR numerical codes, where the flux at the Earth is constant in time.

In the Myriad model, the jelly of sources $q_{\rm acc}$ of the conventional approach is replaced by a constellation of point-like objects located each at $\mathbf{x}_{i}$ and $t_{i}$. The CR flux at the Earth yielded by a population ${\cal P}$ of such a myriad of injectors becomes
\beq
\Phi_{\cal P}(\odot,0) \, = \, {\displaystyle \frac{v_{CR}}{4 \pi}} \;
{\displaystyle \sum_{i \in {\cal P}}} \;
G(\odot , 0 \, \leftarrow \, \mathbf{x}_{i} , t_{i}) \,
q_{\rm acc}(\mathbf{x}_{i} , t_{i}) \equiv
{\displaystyle \sum_{i \in {\cal P}}} \; \varphi_{i} \; .
\eeq
Although we can have some information on the closest and youngest sources, we have little knowledge of the actual population ${\cal P}$ in which we live. This is a problem insofar as the flux $\Phi_{\cal P}$ depends precisely on how the sources are distributed in space and time around us. To tackle the calculation of the flux in these conditions, we can adopt a statistical point of view and consider the ensemble of all possible populations ${\cal P}$ of sources.
Without loss of generality, we may assume that each source accelerates the same CR yield $q_{\rm SN}$, lies within the magnetic halo and is younger than some critical value ${\cal T}$ that sets the size of the phase space volume over which the analysis is carried out. Sources older than ${\cal T}$ have no effect on the flux at the Earth if that age is taken to be a few times the CR confinement time inside the magnetic halo. We also assume a constant explosion rate $\nu \sim 1$ to $3$ per century. Each population contains a number ${\cal N} = \nu \times {\cal T}$ of sources yielding each a flux $\varphi_{i}$ at the Earth whose sum is the flux $\Phi_{\cal P}$.
We may finally assume that each source is {\it independently} and {\it randomly} distributed in phase space according to the probability distribution function (pdf) ${\cal D}(\mathbf{x}_{S},t_{S})$.
In the Myriad model, the individual flux $\varphi$ yielded by a single source is a random variable whose pdf $p(\varphi)$ is the key of the statistical analysis. It is related to the phase space pdf ${\cal D}(\mathbf{x}_{S},t_{S})$ through
\beq
dP \, = \, p(\varphi) \, d{\varphi} \, = \,
{\displaystyle \int_{\displaystyle \, {\cal V}_{\varphi}}} \,
{\cal D}(\mathbf{x}_{S},t_{S}) \; d^{3}\mathbf{x}_{S} \; dt_{S} \; ,
\label{eq:dP}
\eeq
where ${\cal V}_{\varphi}$ denotes the space-time region inside which a source contributes a flux in the range between $\varphi$ and $\varphi + d\varphi$.

\begin{figure}[t!]
\centering
\includegraphics[scale=0.34]{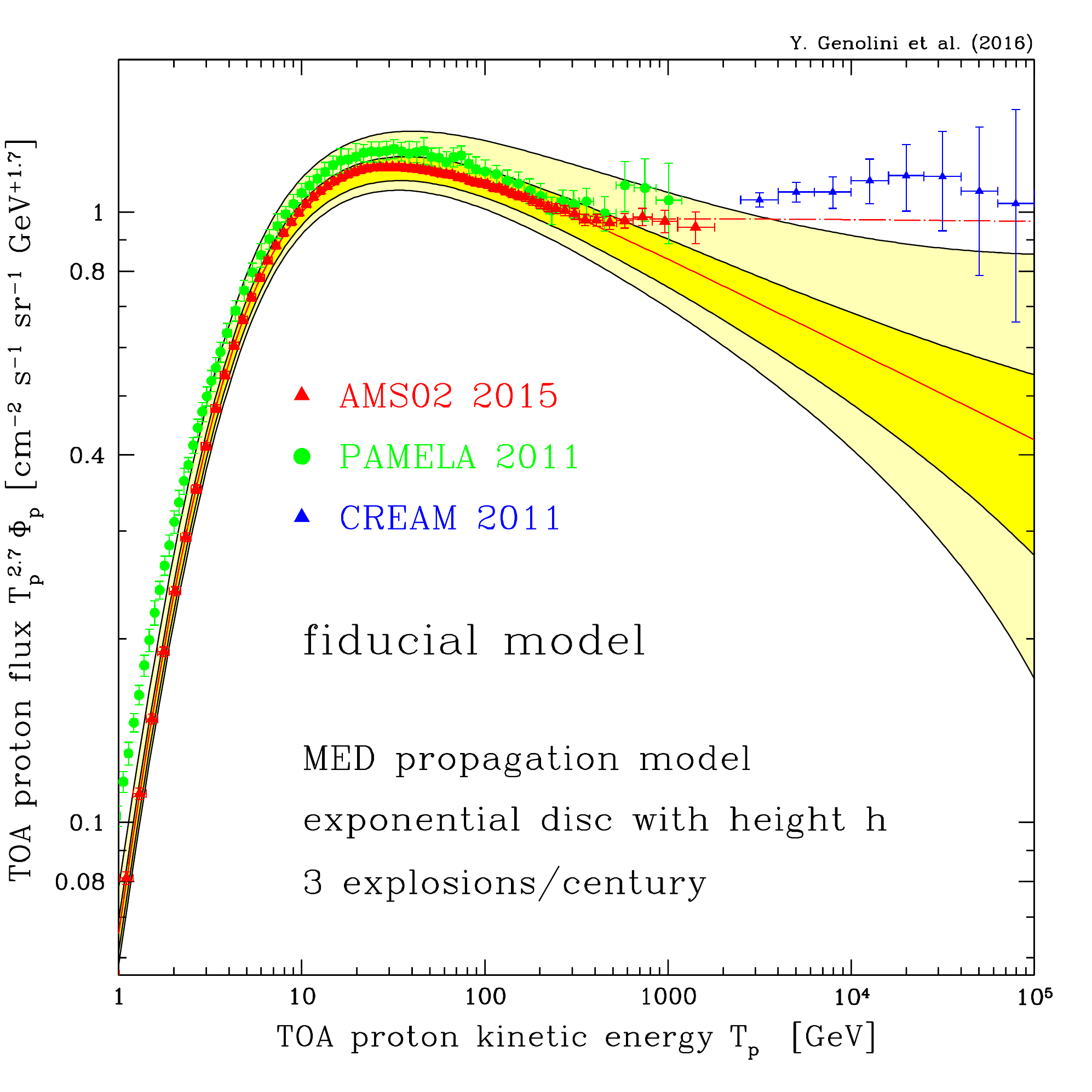}
\includegraphics[scale=0.34]{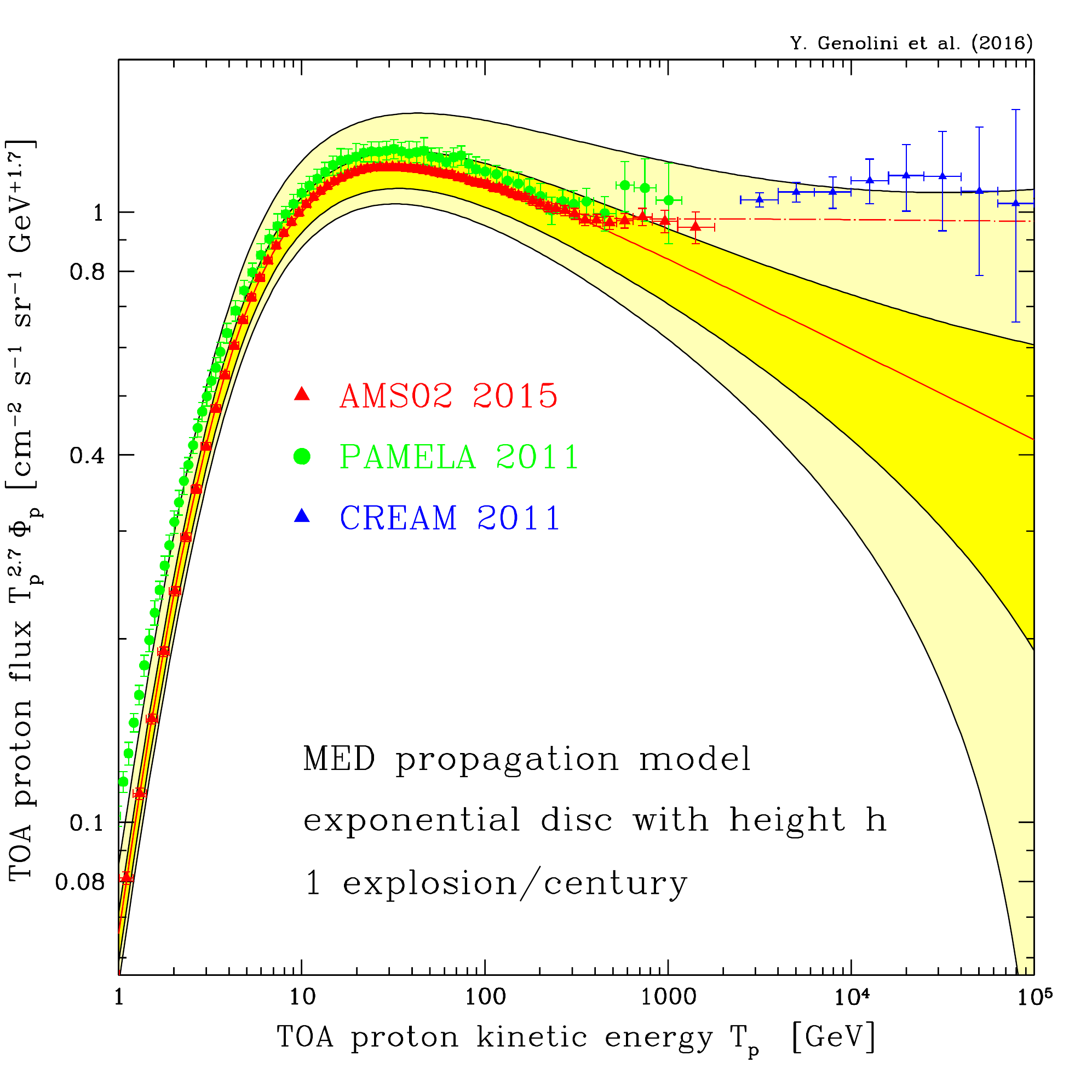}
\vskip -0.25cm
\caption{
In both panels, the TOA proton flux is plotted as a function of TOA proton kinetic energy. The 1-$\sigma$ (pale yellow) and 2-$\sigma$ (dark yellow) bands predictions of the Myriad model have been derived assuming a stable law with index $5/3$. The average flux is featured by the solid red line and is assumed to follow a simple power-law. Actually, we have used here a fit to the AMS-02 data from which the hardening has been removed. Measurements from AMS-02~\cite{2015PhRvL.114q1103A}, PAMELA~\cite{Adriani:2011cu} and CREAM~\cite{2011ApJ...728..122Y} are also shown for comparison. The dotted long-dashed red curve is a fit to the AMS-02 data with hardening taking place at a rigidity of 355 GV. The MED CR propagation model~\cite{Donato:2003xg} has been assumed here, with sources exponentially distributed along the vertical direction with scale height 100~pc. The fiducial case of 3 explosions per century is presented in the left panel. In the right panel, 1 explosion per century is assumed, hence wider uncertainty bands.}
\label{fig:1}
\vskip -0.25cm
\end{figure}

All the conditions are now met to use the central limit theorem in order to derive the pdf of the total flux $\Phi$. Notice that $p(\varphi)$ scales as ${\varphi}^{-8/3}$ in the large flux limit so that the generalized version of the theorem must be applied, with the consequence that the variable $({\Phi} - \langle {\Phi} \rangle)/{\Sigma_{\Phi}}$ is distributed according to the stable law ${\rm S}(5/3,1,1,0;1)$.
Here, $\langle {\Phi} \rangle$ denotes the {\it statistical ensemble} average flux. It is equal to the flux yielded by the smooth distribution of sources $q_{\rm acc}(\mathbf{x}_{S},t_{S}) = {\cal N} \, q_{\rm SN} \, {\cal D}(\mathbf{x}_{S},t_{S})$ of the conventional approach.
The typical flux variance $\Sigma_{\Phi}$ scales like $q_{\rm \, SN} \, K^{-3/5} \, \nu^{3/5}$.
Notice that a similar result was found by~\cite{Mertsch:2010fn} in a pioneering analysis of the CR electron and positron fluxes produced by discrete stochastic sources.

\section{Computing the odds of the Galactic lottery}
\label{sec:results}

We can apply the Myriad model to calculate the pdf of the proton flux at the Earth and estimate the p-value associated to the AMS-02 and CREAM measurements. The 1-$\sigma$ (pale yellow) and 2-$\sigma$ (dark yellow) theoretical uncertainty bands are featured in figure~\ref{fig:1} for two values of the explosion rate. The corresponding p-values are listed in table~\ref{tab:1}.
We can conclude from these results that the observed CR proton and helium anomalies have little chance to originate from a statistical fluctuation in the positions of the local sources. The p-values are larger in the last row of table~\ref{tab:1} where the CR propagation model~A of \cite{Bernard:2012pia} is assumed with a half-height of the magnetic halo of 1.5~kpc and a  diffusion spectral index of 0.85. Both values are now in tension with recent observations.
Notice that the statistical analysis sketched above needs to be refined in two respects: (i) the behavior of the pdf $p(\varphi)$ can be dominated for intermediate values of the flux $\varphi$ by sources distributed along the Galactic plane, hence a stable law with index 4/3 instead of 5/3 and (ii) a light cone cut-off needs to be implemented in relation~(\ref{eq:dP}) since CR diffusion cannot be faster than light. These refinements have been studied in detail in a comprehensive analysis~\cite{Genolini:2016hte} where stable laws are shown to provide a robust description of the statistical behavior of the CR flux.

\begin{table}[h!]
\centering
\caption{The p-values in {\%} of the three last AMS-02 (red) and four first CREAM (blue) proton flux measurements have been calculated in the framework of the Myriad model. They are obtained from the convolution of the experimental uncertainty with a stable law with index 5/3. The three first rows correspond to the MED CR propagation model~\cite{Donato:2003xg} where the SN explosion rate has been decreased from 3 to 1 per century. The last row refers to model A found in~\cite{Bernard:2012pia} to explain the CREAM data as resulting from known local sources.}
\label{tab:1}
{\begin{tabular}{|l|ccccccc|}
\hline
\hline
Kinetic energy [TeV] &
\textcolor{red}{0.724} & \textcolor{red}{0.96} & \textcolor{red}{1.41} &
\textcolor{blue}{3.16} & \textcolor{blue}{5.02} & \textcolor{blue}{7.94} & \textcolor{blue}{12.6} \\
\hline
MED model with $\nu$ = 3 [\%] &
\textcolor{red}{10.2} & \textcolor{red}{8.68} & \textcolor{red}{7.67} &
\textcolor{blue}{1.6} & \textcolor{blue}{1.23} & \textcolor{blue}{1.18} & \textcolor{blue}{0.98} \\
\hline
MED model with $\nu$ = 2 [\%] &
\textcolor{red}{12.3} & \textcolor{red}{10.6} & \textcolor{red}{9.34} &
\textcolor{blue}{2.14} & \textcolor{blue}{1.64} & \textcolor{blue}{1.57} & \textcolor{blue}{1.31} \\
\hline
MED model with $\nu$ = 1 [\%] &
\textcolor{red}{16.2} & \textcolor{red}{14.2} & \textcolor{red}{12.6} &
\textcolor{blue}{3.52} & \textcolor{blue}{2.7} & \textcolor{blue}{2.59} & \textcolor{blue}{2.16} \\
\hline
model A~\cite{Bernard:2012pia} with $\nu$ = 0.8 [\%] &
\textcolor{red}{27.2} & \textcolor{red}{25.8} & \textcolor{red}{24.5} &
\textcolor{blue}{14.3} & \textcolor{blue}{12.8} & \textcolor{blue}{13.3} & \textcolor{blue}{12.9} \\
\hline
\hline
\end{tabular}}
%
\end{table}

\begin{acknowledgement}
P.S. would like to thank the organizers of the 6th RICAP Conference for the friendly and inspiring atmosphere of the meeting. The authors wish to thank Philippe Briand for his kind help on the generalized central limit theorem. This work was supported by Institut Universitaire de France (IUF).
\end{acknowledgement}

%
%
%

%
\vskip 1.0cm
\begin{center}
{\bf
\line(1,0){300}}
\end{center}
%
\end{document}